\documentclass{article}
\usepackage{spconf,amsmath,graphicx} % Use this and comment out two lines above
\usepackage{amssymb,bm}
\usepackage{url,cite}
\usepackage{fancyhdr}

\title{\uppercase{Cascaded all-pass filters with randomized center frequencies and phase polarity for
acoustic and speech measurement and\\ data augmentation}}
\twoauthors
  {Hideki Kawahara}
	{Center for Innovative Research and Liaison\\
	Wakayama University,	Wakayama, Japan}
  {Kohei Yatabe}
	{Department of Intermedia Art and Science\\
	Waseda University,	Tokyo, Japan}
\begin{document}
\ninept

\maketitle
\begin{abstract}
We introduce a new member of TSP (Time Stretched Pulse) for acoustic and speech measurement infrastructure, 
based on a simple all-pass filter and systematic randomization. 
This new infrastructure fundamentally upgrades our previous measurement procedure, 
which enables simultaneous measurement of multiple attributes, including non-linear ones
without requiring extra filtering nor post-processing. 
Our new proposal establishes a theoretically solid, flexible, and extensible foundation in acoustic measurement.
Moreover, it is general enough to provide versatile research tools for other fields, such as biological signal analysis.
We illustrate using acoustic measurements and data augmentation as representative examples among various prospective applications. 
We open-sourced MATLAB implementation. 
It consists of an interactive and real-time acoustic tool, MATLAB functions, and supporting materials. 
\end{abstract}

\thispagestyle{fancy}
\rhead{\scriptsize{~ }}
\lfoot{\footnotesize{\textcopyright ~2021 IEEE Personal use of this material is permitted. Permission from IEEE must be obtained for all other uses, in any current or future media, including reprinting/republishing this material for advertising or promotional purposes, creating new collective works, for resale or redistribution to servers or lists, or reuse of any copyrighted component of this work in other works.}}
\cfoot{ }
\renewcommand{\headrulewidth}{0pt}

\section{Introduction}
\label{ss:intro}
There is a vast knowledge gap in our auditory processing principle connecting waveform and perception, especially in a fine time scale.
The signal family we propose may provide the key to fill the gap.

Phenomena that illustrates the gap are:
A broad class of sounds is perceived identical, although they are entirely different in sampled value levels at the audio sampling rate.
%For example, independent Gaussian white noise having the same variance is a classic example.
Waveform modification by group delay within 0.5~ms is not detectable\cite{brauert1978jasa}.
The detection threshold of a brief noise burst varies 20~dB depending on the burst location inside one pitch period\cite{Skoglund2000ieeetrans}.
%Phase alignment between harmonic components of periodic signals provides distinct timbre differences only for low-pitched signals\cite{patterson1987jasa}.
Group delay compensation of the traveling wave propagation on the basilar membrane enhances the evoked potential response while reduces ``compactness'' of the unit pulse, and the unit pulse without group delay modification yields the most ``compactness''\cite{UPPENKAMP200171}.
These phenomena motivated the first author to discard the phase information of speech sounds in STRAIGHT\cite{kawahara1999spcom}.
The invention of FVN (Frequency domain variant of Velvet Noise)\cite{kawahara2018interspeech} is the most recent example driven by this motivation.
%We designed FVN as the excitation source of synthetic speech
We designed it for the excitation source of synthetic speech.

Accidentally, we found that FVN provides powerful tools for assessing acoustic systems\cite{kawahara2020simultaneous}. %, which consist of non-linearity and temporal variability\cite{kawahara2020simultaneous}.
FVN is a new member of TSP (Time Stretched Pulse)\cite{schroeder1979integrated,aoshima1981jasa,muller2001transfer,ochiai2013a,guidorzi2015impulse,burrascano2019swept} in a broad sense.
Well-known TSP members are pseudo-random noise (PN) and swept-sine (SS) signals.
They provide an essential infrastructure of acoustic measurement and assessment\cite{schroeder1979integrated,aoshima1981jasa,dunn1993distortion,stan2002comparison,burrascano2019swept,farina2000simultaneous}.
It is crucial to acquire reusable speech materials\cite{svec2010ajsp,Rita2018ajsp} and present sound stimuli to participants of subjective tests using adequately prepared sound devices and the environment.
FVN also added new functionality to assessment tools\cite{kawahara2020simultaneous}.

Concerning the gap, FVN provides a means to augment acoustic data for end-to-end DL-based applications\cite{DLforAudio}.
Systematic group delay modification of the original signal can generate many perceptually identical signals. 
Eventually, the end-to-end approach may yield latent variables that embody still mysterious human auditory processing principles.
If it would provide a clue for uncovering the mystery, it is a desirable consequence of our original motivation lead.

Unfortunately, FVN is not theoretically solid enough to support rigorous scientific and engineering research.
FVN is a mixture of many heuristics that started from time and frequency axes exchange of the original velvet noise\cite{jarvelainen2007reverberation,valimaki2013ieetr}.
FVN design procedure modifies the phase of the transfer function of a unit pulse using a weighted sum of six-term cosine series\cite{kawahara2017interspeechGS}. 
This procedure has many tuning parameters without solid theoretical backgrounds. 
The procedure modifies the phase to manipulate the actual manipulation target, the group delay.
The assignment rule of phase manipulation functions, a modified copy of the original velvet noise, may not be the best.

%We introduced a simultaneous measurement procedure of the linear time-invariant response, 
%the nonlinear time-invariant response, and random and time-varying components\cite{kawahara2020simultaneous}.
%This article proposes the complete replacement of its essential infrastructure, 
%an extended TSP (Time Stretched Pulse) called FVN (Frequency domain variants of Velvet Noise\cite{kawahara2018interspeech}),
%with cascaded all-pass filters with randomized center frequencies and phase polarities (CAPRICEP\footnote{
%Cascaded All-Pass filters with RandomIzed CEnter frequencies and Phase polarities}).

%=== glue paragraph will be placed here ===
This article proposes another member of TSP called CAPRICEP\footnote{
Cascaded All-Pass filters with RandomIzed CEnter frequencies and Phase polarities}.
%It replaces FVN and provides a theoretically sound foundation. 
CAPRICEP removes all these difficulties and expands FVN's functionality.
IIR all-pass filter, a starting building block of CAPRICEP, has only one tuning parameter, the root's proximity to the unit circle.
IIR all-pass filter has a unique phase function shape.
It provides means to directly design the group delay instead of indirectly design using the phase.
Built-in numerical optimization assures that the generated CAPRICEP is optimum.

The significant contribution of this article is the proposal of CAPRICEP.
The following section provides the derivation of CAPRICEP
followed by a brief introduction of the simultaneous measurement method of multiple responses.
The derivation of this simultaneous measurement method is a replication of the FVN-based method's\cite{kawahara2020simultaneous}. 
CAPRICEP adds flexibility and extended functionality because of its building block, all-pass filter.

Then, we introduce two representative examples to illustrate the possibilities of CAPRICEP-based methods.
First is an interactive and real-time acoustic measurement tool.
The tool is for measuring sound acquisition and sound reproduction systems. 
Proper assessment based on these measurements is crucial for 
preparing and evaluating sound materials. 
The second example concerns the knowledge gap. 
Using a set of unit-CAPRICEPs as a filter-bank, it provides perceptually indistinguishable sounds from original recordings. 
It works as a data augmentation device for the end-to-end approach\cite{DLforAudio}.
Finally, we review other related works and other possible applications.
We found that CAPRICEP enables a new approach for investigating 
the role of auditory perception while speaking and singing\cite{kawahara1994interactions,Kawahara1996is,TOURVILLE20081429,houde2013PNAS,Zarate2013front}.

%\section{Background}
%\label{ss:background}
%Short noise bursts modify spectral shapes when used in speech synthesis because of the statistical fluctuation. 
%This modification is not desirable. 
%We need noise bursts that do not modify the spectral shape. 
%An excitation source signal for speech synthesis, which can traverse from periodic signal to random signal seamlessly, 
%is beneficial for implementing mixed-mode excitation. 
%We proposed FVN as a solution to these requirements\cite{kawahara2018interspeech}.

%FVN is a member of TSP. 
%We realized that FVN has useful features that existing TSP members do not have. 
%TSP signals play essential roles in measuring the system's response characteristics. 
%FVN enables the simultaneous measurement of various system attributes without introducing extra equipment and post-processing\cite{kawahara2020simultaneous}.

%\section{Goal of the TSP design}
%\label{ss:designGoal}
%We want to design a TSP which has the following attributes:
%\begin{enumerate}
%    \item The waveform is temporally localized.
%    \item It has many degrees of freedom to design to generate items.
%    \item The generated items are close to orthogonal each other.
%\end{enumerate}

\section{Proposed TSP: CAPRICEP}
An all-pass filter is a filter whose gain transfer function is a frequency-independent constant. 
Other than the trivial solution, a unit impulse, an all-pass filter has an impulse response that temporally spreads. 
It is a member of TSP signals, in a broad sense. 
Cascaded all-pass filters also yield an all-pass filter. 
By introducing a systematic procedure, we can design useful attributes to the yielded filter's impulse response. 

The design goal of the response is as follows:
a) The waveform is temporally localized, b) It has many degrees of freedom in designing the waveform,
and consequently c) Cross-correlation between different generated waveform is close to zero.
In other words, we want noise bursts with a completely flat (frequency-independent) power spectrum.
CAPRICEP fulfilled the goal. %, as summarized at the end of this section.

\subsection{Building block: all-pass filter}
\label{ss:buildingBlock}
Let us start from its building block. 
The following equation defines the transfer function of a simple discrete-time all-pass filter, 
characterized by the center frequency $f_k$ and the bandwidth $b_k$\cite{oppenheimerBook}. 
\begin{align}
    %H_k(z) & = \frac{(z - z_k^\ast) (z - z_k)}{(1 - z_k z)(1 - z_k^\ast z)} \\
    %z_k & = r_k\exp(j\theta_k) \nonumber \\
    %r_k & = \exp\left(-\frac{\pi b_k}{f_s}\right), \ \ \theta_k = \frac{2\pi f_k}{f_s} , \nonumber
    %z_k & = \exp\left(-\frac{\pi b_k}{f_s} + j \frac{2\pi f_k}{f_s} \right), \nonumber
    H_k(z) & = \frac{z^{-1} - z_k^{\ast}}{1 - z_k^{\ast} z^{-1}} , \ \ \ \ 
    \left(\mbox{Note:} \ \ H_k^{\ast}(z) H_k(z) = 1\right) \\
      z_k & = \exp\left(-\frac{\pi b_k}{f_s} + j \frac{2\pi f_k}{f_s} \right), \label{eq:allPParam}
\end{align}
where $f_s$ represents the sampling frequency and $j = \sqrt{-1}$.
A term $a^\ast$ represents the complex conjugate of $a$.
By cascading these building blocks also yields an all-pass filter.
Its inverse z-transform provides an impulse response.
We call the impulse response unit-CAPRICEP afterward.

\subsection{Cascaded all-pass filters: CAPRICEP}
\label{ss:cascading}
The transfer function of the cascaded all-pass filters is the product of constituent transfer function $H_k(z)$.
\begin{align}
    H_C(z) & = \prod_{k = 1}^K H_k(z) ,
\end{align}
where $K$ represents the total number of the cascaded filters.
A systematic procedure determines each center frequency $f_k$
using two set of random numbers $r_1[k]$ and $r_2[k]$.
We introduce random numbers because each pole corresponds to a damped sinusoid.
Intuitively speaking, the sum of many sinusoids with the same amplitude with the random phase yields items behave like samples from normal distribution\cite{lyon1970jasa}.
We hope frequency randomization yields noise-like behavior of the impulse response of $H_C(z)$.

%An all-pass filter introduces a peak group delay at $f_k$.
%The sum of many group delays results in a considerable group delay.
%We do not want to temporally shift the processed waveform (at least for the first time).
%We use the second random number to flip the response time axis with a 50\% probability.

%These considerations led to the following equation:
The following equation is an implementation of the systematic procedure of filter assignment.
\begin{align}\label{eq:fkdesign}
    f_k & = r_1[k] \,F_d \sum_{k = 1}^K r_2[k] ,
\end{align}
where $F_d$  represents the average interval between aligned center frequencies (\textit{i.e.} $f_{k+1} > f_k$).
The random number $r_1[k]$ has the value $1$ and $-1$ with equal probability.
Note that $r_1[k] = -1$ makes $H_k(z)$ anti-causal and makes its impulse response time-reversed.
This randomization is for design flexibility described in the next section.
Without introducing $r_1[k]$, the response is causal and does not provide flexible design.

The random number $r_2[k]$ obeys a probability density distribution function (pdf) $g(x)$ defined on $[0, 1]$.
We do not randomize $f_k$ directly.
Instead, we randomized the frequency interval between neighboring all-pass filters.
References\cite{jarvelainen2007reverberation,valimaki2013ieetr} reported simple random allocation does not yield well-behaved velvet noise\footnote{
They named the noise ``velvet noise'' because the noise sounds smoother than Gaussian white noise. Simple random allocation destroyed the smoothness\cite{jarvelainen2007reverberation,valimaki2013ieetr}.
}.
Intuitively speaking, the original velvet noise and FVN use the triangular pdf for interval distribution.
They also use the additional constraint, fixed-width partitioning of the time axes. %s (original velvet noise), and the frequency axis (FVN).
%We observed that the triangular pdf and fixed-width partitioning caused glitches in FVN.
This fixed-width partitioning forced us to introduce non-linear frequency axis warping for shaping temporal distribution\cite{kawahara2018interspeech}.

We introduce the Beta distribution for $r_2[k]$ because it is a simple adjustable distribution having two design parameters.
%Let name the pdf $b(x)$. 
The following is the definition of the Beta distribution.
\begin{align}
    b(x) & = \frac{x^{\alpha - 1}(1 - x)^{\beta - 1} \Gamma(\alpha + \beta)}{\Gamma(\alpha) \Gamma(\beta)} ,
\end{align}
where $\Gamma(\alpha)$ is Gamma function.
When $\alpha \equiv \beta$, $b(x)$ seamlessly deforms from concave to convex via uniform distribution with $\alpha$.

For $b_k$ (Eq.~\ref{eq:allPParam}), we make it %independent of $k$ and 
proportional to $F_d$ using a coefficient $c_\mathrm{mag}$.
In short, we set $b_k = c_\mathrm{mag}F_d$ (at least for the first time).
In summary, $H_C(z)$ has three design parameters, $F_d, c_\mathrm{mag}$, and $\alpha$.
This additional parameter $\alpha$ that FVN does not have enables us to design the shape of the waveform distribution.
Let $h_c[n]$ represent the impulse response of $H_C(z)$.
It is unit-CAPRICEP.

\subsection{Shaping waveform distribution of unit-CAPRICEP}
\label{ss:shapingDesign}
%We want to design the variance $V(h_c[n])$ of the instantaneous value of the impulse response of $H_C(z)$ on the discrete-time axis $n$ to have
%a specific target shape $W[n]$
Shaping waveform distribution is an optimization problem.
Let us define the target shape on the discrete-time axis $W[n]$.
%We want to make the variance $V(h_c[n])$ have the shape $W[n]$.
We design the sample variance $V(h_c[n])$ by minimizing a distance measure between $V(h_c[n])$ and $W[n]$.
We use Wasserstein distance\cite{vallender1974siam} %between $V(h_c[n])$ and $W[n]$ 
as the cost function to minimize because it is an energy allocation problem on the discrete-time axis, in the other words.
%Implementation of this optimization uses IFFT (inverse fast Fourier transform) for calculating the impulse response, and uses the sample variance $V_s(h_c[n])$ instead of the variance $V(h_c[n])$.
%\begin{align}
%    V_s(h_c[n]) & = \frac{1}{M-1} \sum_{m = 1}^M \left| h_c^{(m)}[n] \right|^2,
%\end{align}
%where $h_c^{(m)}[n]$ represents the $m$-th impulse response
%and $M$ represents the total number of the generated impulse responses.

We selected the rectangular shape as the first target because it is suitable for making less-fluctuating test signals using periodic allocation (\textit{i.e.} OLA: overlap and add) of unit-CAPRICEPs.
%We defined cost function by setting $F_d$ a fixed parameter % and setting $c_\mathrm{mag} = 2^{\frac{1}{4}}$ and $\alpha = 8$.
%and optimized the design parameters $c_\mathrm{mag}$ and $\alpha$ using a coarse grid search. 
%Then, we 
We conducted several coarse multivariate optimizations before final optimization.
It gave us the setting $c_\mathrm{mag} = 2^{\frac{1}{4}}$ and $\alpha = 8$.
Then, we optimized the duration of the rectangle $T_\mathrm{ERD}$ using a grid search.
\begin{figure}
    \centering
    \includegraphics[width=0.48\hsize]{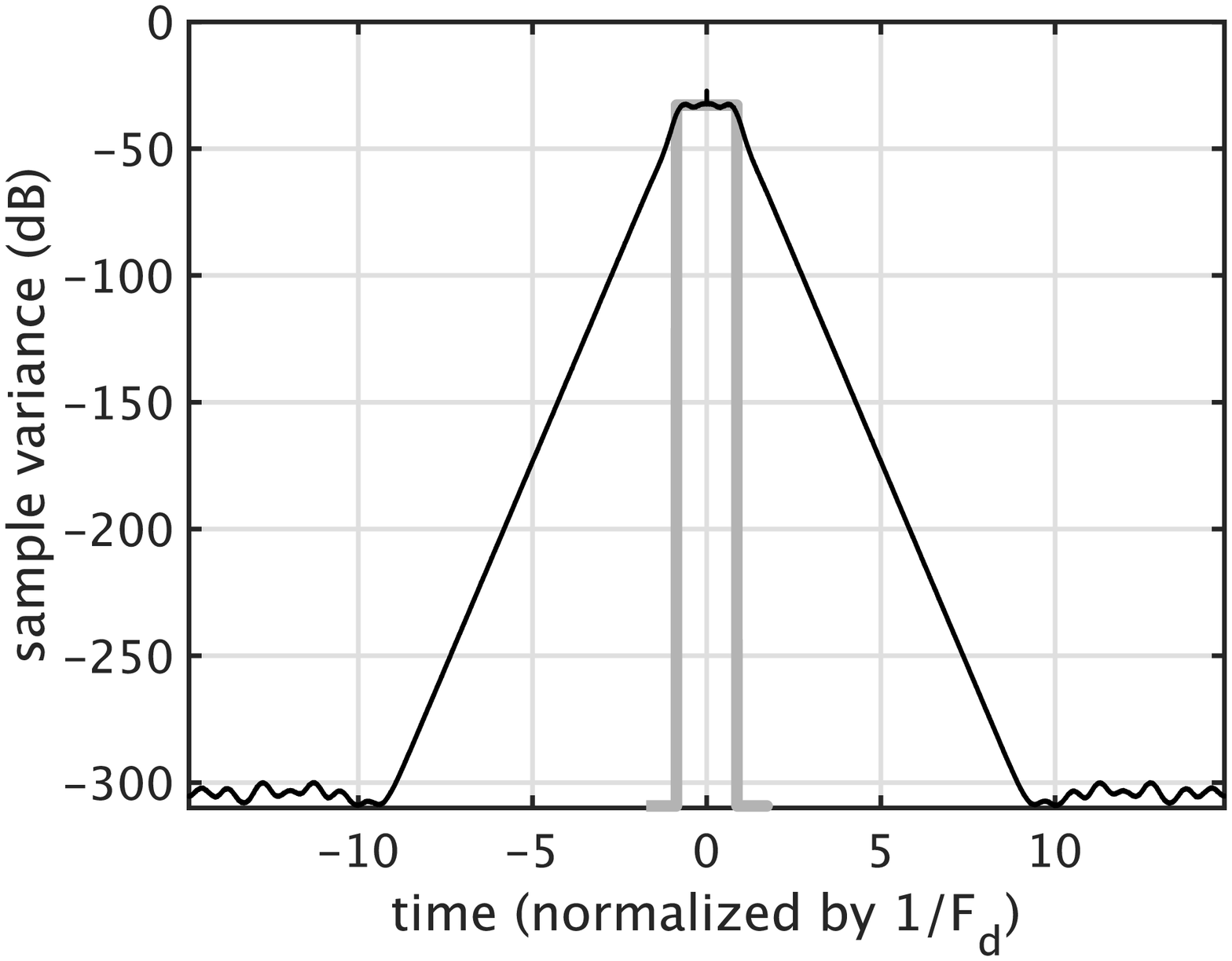}
    \hfill
    \includegraphics[width=0.48\hsize]{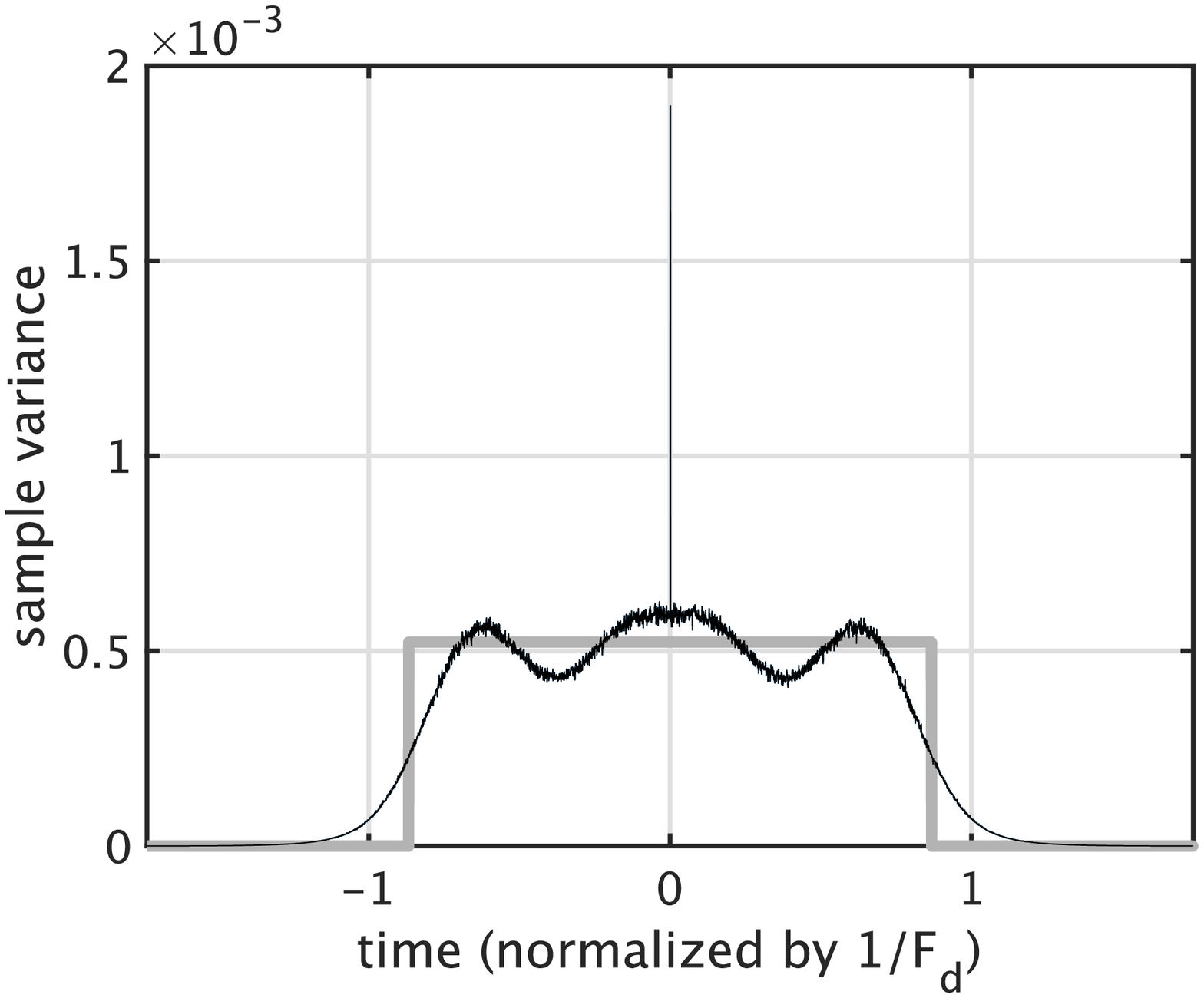}
    \vspace{-3mm}
    \caption{Example of shape design.
    The black line shows the sample variance. The gray box shows the rectangle of the same total energy.}
    \label{fig:shapeDesign}
\end{figure}

Figure~\ref{fig:shapeDesign} shows an example of shape design.
The sample variance calculation used 5000 unit-CAPRICEPs. % with $f_s=44100$~Hz, $F_d=40$~Hz.
The optimized $T_\mathrm{ERD}$ is 1.736 times of the nominal duration of $1/F_d$.
Note that the decay outside of the rectangle is exponential.
%This steep decay satisfies the first design goal in Section~\ref{ss:designGoal}.
The total number of cascaded filters is 1102 for $f_s=44100$~Hz, $F_d=40$~Hz. % and satisfies the second design goal.
%The outstanding pulse at the origin does not become harmful because applying another CAPRICEP with a short duration (several ms) removes it.

\begin{figure}
    \centering
    \includegraphics[width=0.45\hsize]{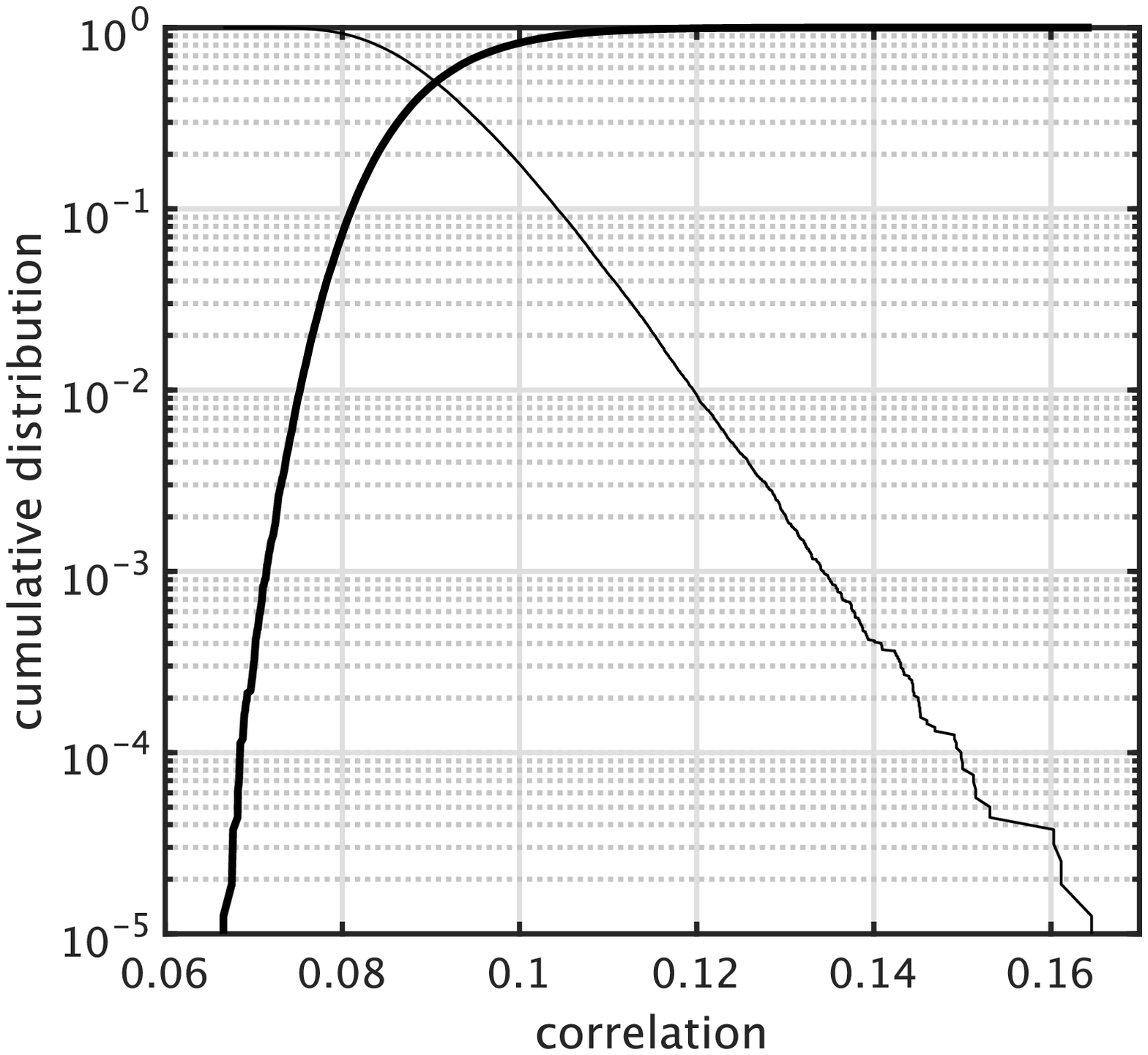}
    \hfill
    \includegraphics[width=0.51\hsize]{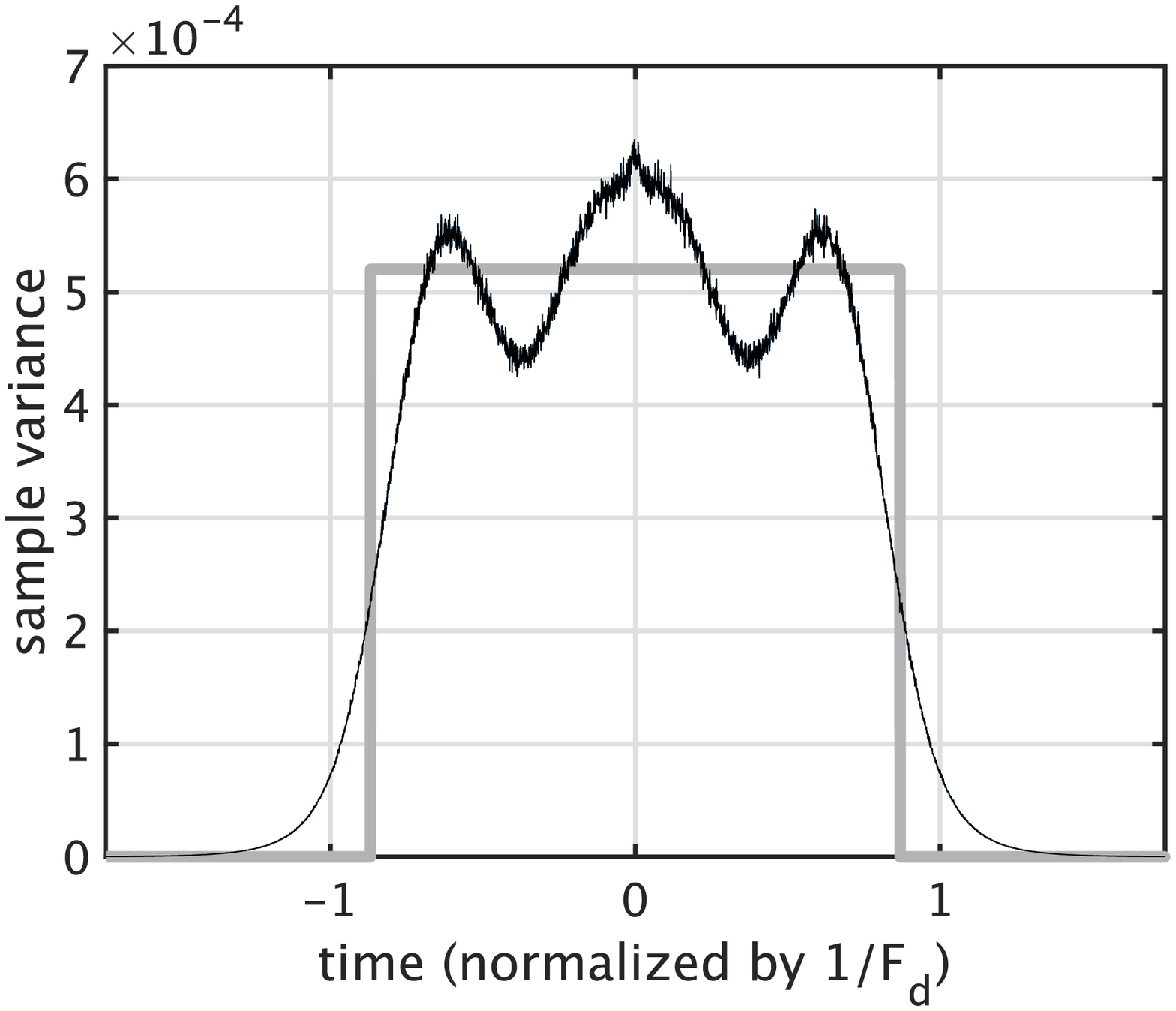}
    \vspace{-3mm}
    \caption{(Left) Distribution of maximum cross correlation absolute values of
    $400 \times 399 $ pairs of different CAPRICEPs.
    (Right) Composite design example of a unit-CAPRICEP.}
    \label{fig:crossCorr}
\end{figure}
Left plot of Fig.~\ref{fig:crossCorr} shows the distribution of the maximum cross correlation (absolute value) of $400 \times 399 $ pairs of different unit-CAPRICEPs.
The thick line shows probability $p(x_c | \ \ |x_c| \le \theta_c)$ where $x_c$ represents the maximum cross correlation and $\theta_c$ represents the threshold correlation value shown in the horizontal axis.
The thin line shows $p(x_c | \ \ |x_c| > \theta_c)$.
The median of the distribution is 0.0905 and close to zero.
%This small correlation satisfies the third design goal.

Another useful target shape is the raised cosine because
OLA with 50\% overlap provides a constant level.
Cascading a short (0.5~ms) raised cosine-shaped unit-CAPRICEP and the rectangular one yielded the shape shown in the right plot of Fig.~\ref{fig:crossCorr}.
The outstanding spike at the center shown in Fig.~\ref{fig:shapeDesign} disappears.
This composite shape is appropriate for acoustic measurement applications.
%\begin{figure}
%    \centering
%    \includegraphics[width=0.75\hsize]{fixedShapingLinN.eps}
%    \caption{Composite design example of a unit-CAPRICEP.}
%    \label{fig:fixedShapingLinN}
%\end{figure}

\subsection{Summary of this section: goal and the result}
CAPRICEP fulfilled the design goal as follows:
a) Using IIR all-pass filter as the building block made the decaying exponential that enables well-behaving localization (section~\ref{ss:buildingBlock}).
b) Cascading many filters provides many degrees of freedom (section~\ref{ss:cascading} and~\ref{ss:shapingDesign}).
c) Many degrees of freedom made cross-correlation between different unit-CAPRICEPs close to zero (section~\ref{ss:shapingDesign}).

Note that making $T_\mathrm{ERD}$ frequency-dependent is straightforward for CAPRICEP, while complicated and indirect for FVN\cite{kawahara2018interspeech}.
There is much additional flexibility in design using $\beta $ and $b_k$, which are fix-valued in the current implementation.

\section{Simultaneous measurement}
Systematic allocation of unit-FVNs enabled simultaneous measurement of
the linear time-invariant, the non-linear time-invariant responses and
random and time-varying components\cite{kawahara2020simultaneous}.
Replacing FVN with CAPRICEP also enables a similar simultaneous measurement.
The following is a brief outline of the procedure in\cite{kawahara2020simultaneous}.
%Please refer to \cite{kawahara2020simultaneous} for details.

\subsection{Four CAPRICEP sequences and response measurement}\label{ss:fvnMes4seq}
A set of four periodic sequences made from unit-CAPRICEPs enables the simultaneous measurement.
Define a weight matrix $\mathrm{B}_4$, which consists of four orthogonal binary rows shown below.
\begin{align}
\mathrm{B}_4 & = \left[ \begin{array}{rrrrrrrr}
1 & 1 & 1 & 1 & 1 & 1 & 1 & 1 \cr
1 & -1 & 1 & -1 & 1 & -1 & 1 & -1 \cr
1 & 1 & -1 & -1 & 1 & 1 & -1 & -1 \cr
1 & 1 & 1 & 1 & -1 & -1 & -1 & -1
\end{array}
\right] \nonumber \\
& = [\mathbf{b}^{(1)}  \mathbf{b}^{(2)} \mathbf{b}^{(3)} \mathbf{b}^{(4)}]^T . \label{eq:orthMat}
\end{align}

The following equation defines the $m$-th CAPRICEP sequence $s_4^{(m)}[n],$  $(m = 1, \ldots, 4)$.
This procedure is a periodic OLA.
\begin{align}
s_4^{(m)}[n] & = \sum_{k=0}^{K -1} h_c^{(m)}[n - kn_\mathrm{o}]\,b^{(m)}_{\mathrm{mod}(k, 8) + 1} \ ,
\end{align}
where $b_j^{(m)}$ represents the $j$-th element of $\mathbf{b}^{(m)}$.
$K$ represents the number of repetitions.
Note that 8 is the number of columns of $\mathrm{B}_4$.

\subsection{Pulse recovery}
Make a test $x_\mathrm{testR}[n]$ by adding the first three CAPRICEP sequences.
Convolution of the test signals and the time-reversed unit-CAPRICEP yields compressed signal 
$q^{(m)}_\mathrm{R}[n]$.
They consist of recovered pulses corresponding to repetitions.
Since a set of unit-CAPRICEPs are not strictly orthogonal, correlations remain. % as background noise.
%\vspace{1mm}
\begin{align}\label{eq:compress}
q^{(m)}_\mathrm{R}[n] & = h_\mathrm{fvn}^{(m)}[-n] \ast x_\mathrm{testR}[n] ,
\end{align}
where ``$\ast$'' represents the convolution operation.

\subsection{Orthogonalization}
The inner product of $\mathbf{b}^{(i)}$ and $\mathbf{b}^{(j)}$ of (\ref{eq:orthMat}) is zero when $i\ne j$.
Periodic time shift and averaging using the binary weights in $\mathbf{b}^{(1)}, \ldots, \mathbf{b}^{(4)}$ remove these cross-correlations.
The following equation yields the orthogonalized signal $r_\mathrm{itrR}^{(m)}[n]$.
\begin{align}\label{eq:orthogonal}
%r_\mathrm{itr}^{(m)}[n] & = 2^{-M}\sum_{k=0}^{2^{M} - 1} q^{(m)}[n - kT_\mathrm{o}]b^{(m)}_{P(k, M)} ,
%r_\mathrm{itr}^{(m)}[n] & = 2^{-M}\sum_{k=0}^{2^{M-1} - 1} q^{(m)}[n - kT_\mathrm{o}]b^{(m)}_{\widetilde{k+1}} \ ,
r_\mathrm{itrR}^{(m)}[n] & = \frac{1}{8}\sum_{k=0}^{8 - 1} q^{(m)}_\mathrm{R}[n - kn_\mathrm{o}]\,b^{(m)}_{\widetilde{k+1}} \ ,
\end{align}
where $8$ is the length of the weight vectors $\mathbf{b}^{(1)}, \ldots \mathbf{b}^{(4)}$.
The notation $\widetilde{k+1}$ represents cyclic indexing between 1 and 8.

\subsection{Synchronous averaging for unit impulse recovery}
Synchronous averaging given below yields the unit impulse response.
%The following synchronous averaging 
It also reduces the effect of random fluctuation of the signal.
\begin{align}\label{eq:syncAv}
r^{(m)}_\mathrm{R}[n] & = \frac{1}{\#(\Omega)} \sum_{n_\mathrm{ini} + 8 k n_\mathrm{o} \in \Omega} r_\mathrm{itrR}^{(m)}[n + n_\mathrm{ini} + 8 k n_\mathrm{o}] ,
\end{align}
where $0 \le n < n_\mathrm{o}$, and the symbol $\Omega$ represents the region where cross-correlation is effectively vanished.
The function $\#(\Omega)$ yields the number of pulses separated by $8n_\mathrm{o}$ and located inside the region.
Then, further averaging the first three responses $r^{(m)}_\mathrm{R}[n], (m = 1, 2, 3)$ provides the final averaged response $r_\mathrm{R}[n]$.
\begin{align}
r_\mathrm{R}[n] & = \frac{1}{3} \left(r^{(1)}_\mathrm{R}[n] + r^{(2)}_\mathrm{R}[n] + r^{(3)}_\mathrm{R}[n] \right) .
\end{align}

\begin{figure}
    \centering
    \includegraphics[width=0.8\hsize]{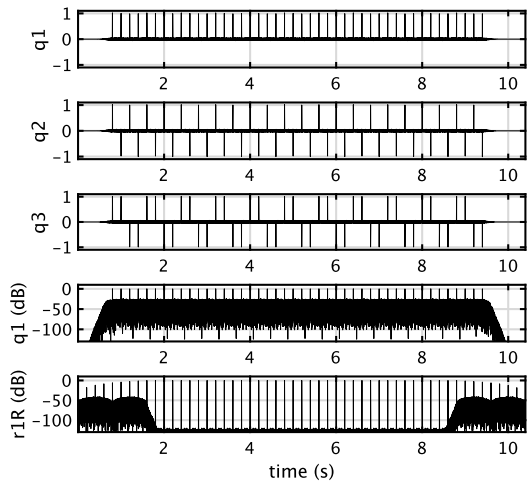}
    \vspace{-3mm}
    \caption{Pulse recovered signals $q_\mathrm{R}^{(m)}[n],$ 
    ($m = 1, 2, 3) $ and orthogonalized signal $r^{(1)}_\mathrm{R}[n]$.
    The fourth plot shows $q_\mathrm{R}^{(1)}[n]$ using dB scale.}
    \label{fig:orthsignals}
\end{figure}
Figure~\ref{fig:orthsignals} shows examples to illustrates these procedures. %\footnote{
This example uses CAPRICEP designed to have the equivalent rectangular duration $T_\mathrm{ERD}=0.2$~s.
We truncated the length of CAPRICEP 0.8~s for implementation
using FFT-based fast convolution.
IIR-based implementation requires a two-pass procedure
consisting of time axis reversal, which is not relevant for
real-time applications.
% because the assumed application, acoustic measurement, rarely has a dynamic range exceeding 90~dB.
%The background level of the bottom plot of Fig.~\ref{fig:orthsignals} is the result of this truncation.
%}.
The top three plots of Fig.~\ref{fig:orthsignals} show the recovered signals $q_\mathrm{R}^{(m)}[n],$ 
    ($m = 1, 2, 3) $.
The fourth plot shows $q_\mathrm{R}^{(1)}[n]$ using dB scale.
The background noise around 45~dB is the result of the cross-correlation between unit-CAPRICEPs.
The bottom plot shows the orthogonalized signal $r^{(1)}_\mathrm{R}[n]$ using the dB scale.
Note that the orthogonalization procedure removes components due to cross-correlation between unit-CAPRICEPs.
%These results replicate that of FVN's results~\cite{kawahara2020simultaneous}.

\vspace{-4pt}
\subsection{Measuring deviations from linear time-invariant systems}\label{ss:decomposition}
The orthogonalized signal using the fourth unit-CAPRICEP $r_\mathrm{itrR}^{(4)}[n]$ does not consist of components in the test signal $x_\mathrm{testR}[n]$.
Therefore, $r_\mathrm{itrR}^{(4)}[n]$ consists of the background noise, the effects of temporal variation in the response, and randomly generated non-linear time-variant response.

The consecutive eight repetition segments provide the eight possible combinations of three unit-CAPRICEPs' polarities.
The averaged response $r_\mathrm{R}[n]$ is the average of all these combinations, deviations from this average consists of the non-linear time-invariant responses (and the random component mentioned above).
Detailed derivation in~\cite{kawahara2020simultaneous} provides the procedure of the simultaneous measurement and simulation tests quantitatively validated the procedure using FVN.
This derivation holds for CAPRICEP because FVN and CAPRICEP are members of TSP.
We conducted the same simulation by replacing FVNs with CAPRICEPs.
The results using CAPRICEP also validated the procedure.

\vspace{-4pt}
\section{Application and discussion}
\vspace{-2pt}
We introduce two examples of CAPRICEP's application,
an interactive acoustic measurement tool and data augmentation.

\begin{figure}
    \centering
    \includegraphics[width=0.9\hsize]{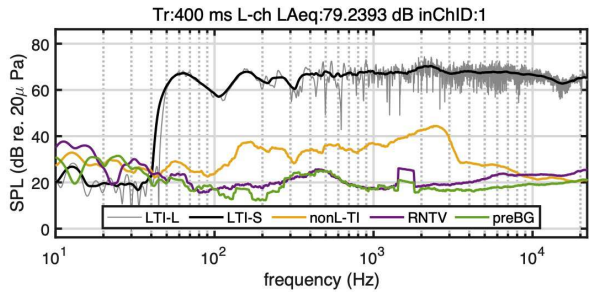}\\
    \vspace{-4mm}
    \caption{Simultaneous measurement of multiple acoustic attributes.
    Top two lines represents the linear time-invariant response (raw: LTI-L and smoothed: LTI-S). 
    The middle line represents the non-linear time-invariant response (nonl-TI), and the
    bottom two lines represent the background noise (pre-BG) and
    random and time-varying response (RNTV).
    Refer\cite{kawahara2020simultaneous} for details.}
    \label{fig:multipleMes}
\end{figure}
Figure~\ref{fig:multipleMes} shows an example of simultaneous multiple attribute measurement. 
Distance between the microphone and the powered loudspeaker was 50~cm
located in a home office.
We used an interactive and real-time acoustic measurement tool\cite{kawahara2021capricep} based on CAPRICEP, which is a
substantial upgrade of our FVN-based tool\cite{kawahara2020simultaneous}.

%\subsection{Interactive and real-time tool}

%\subsection{Set of unit-CAPRICEPs}
\begin{figure}
    \centering
    \includegraphics[width=0.45\hsize]{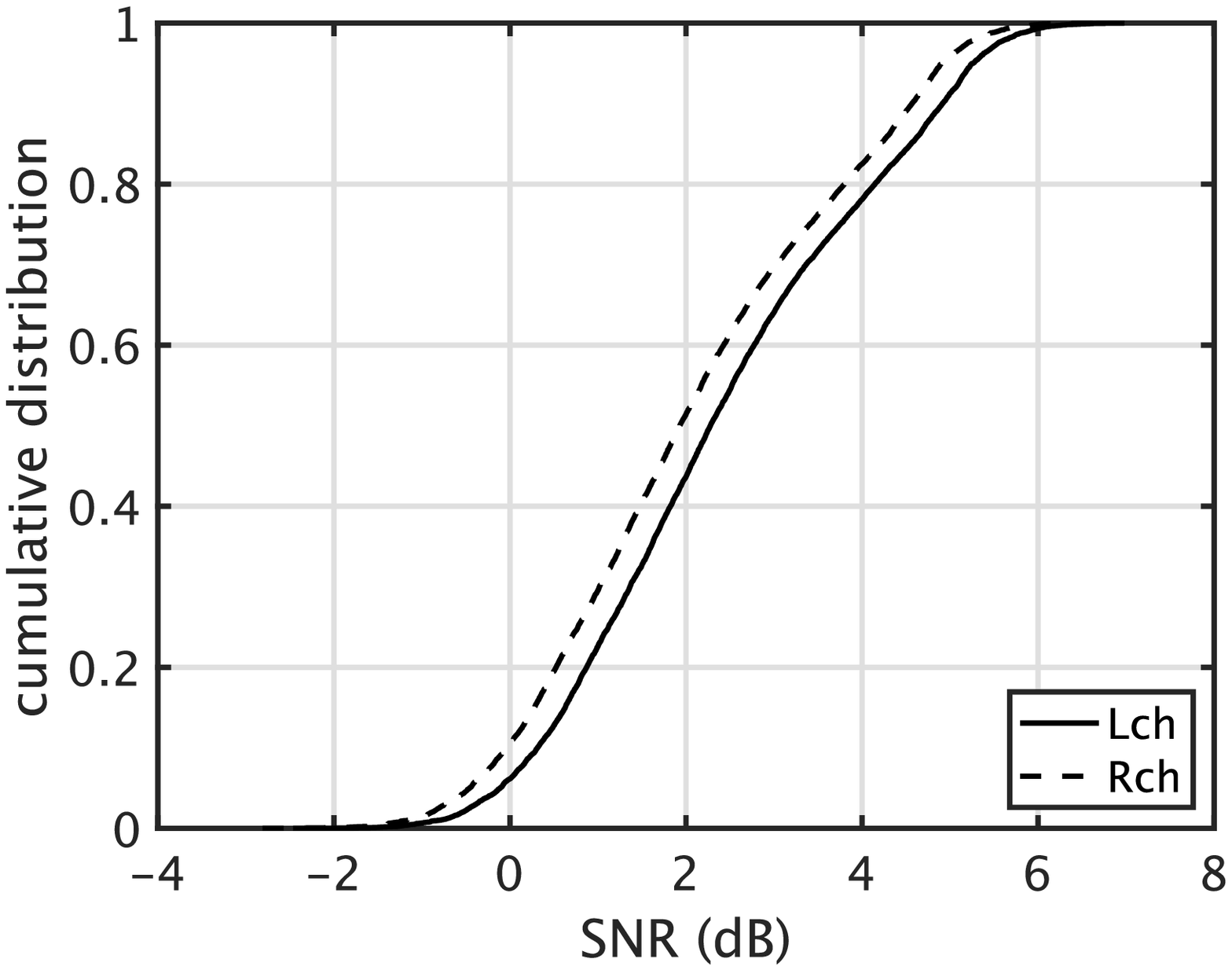}
    \hfill
    \includegraphics[width=0.50\hsize]{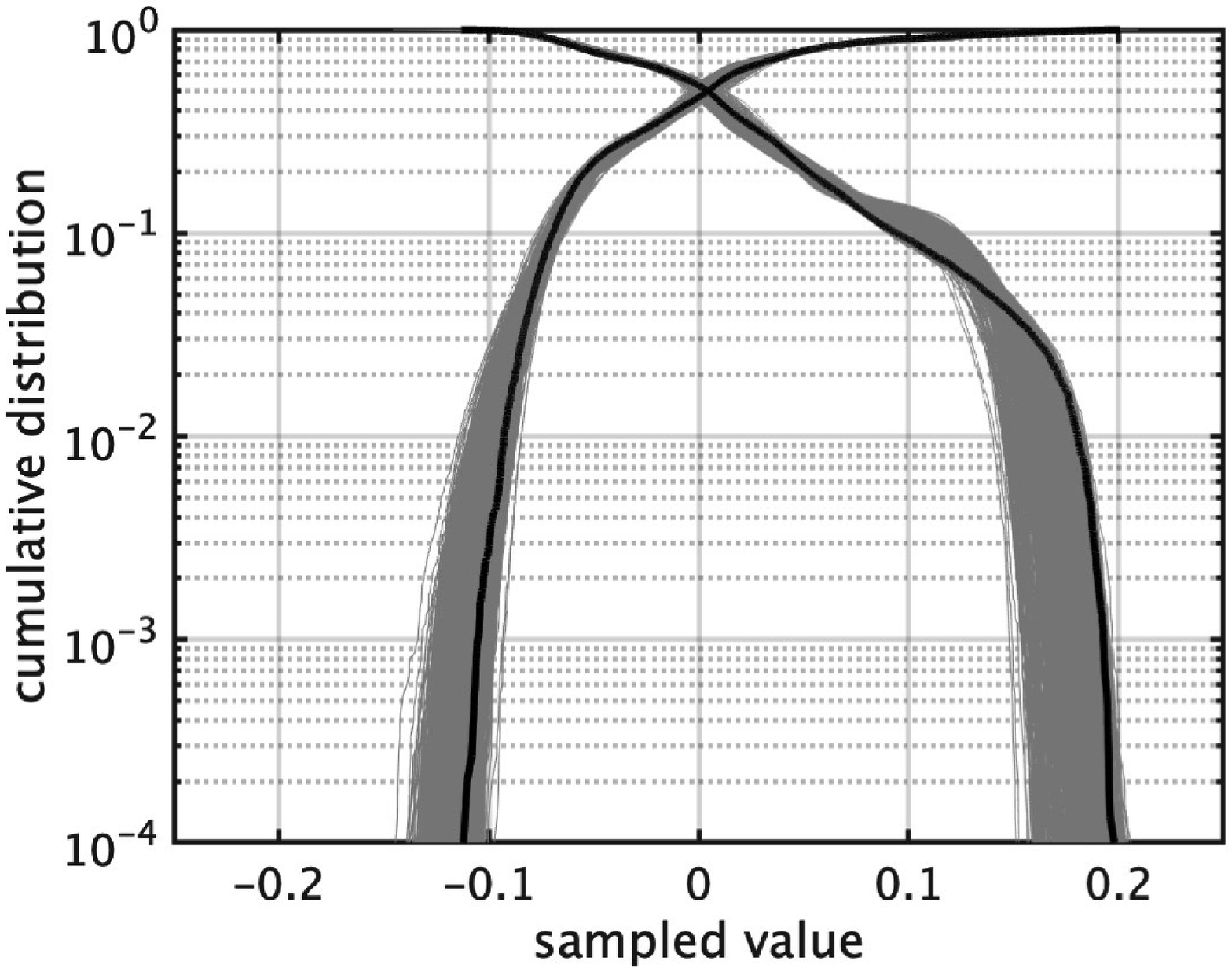}
    \vspace{-3.5mm}
    \caption{Effects of waveform modification using unit-CAPRICEPs for filtering. 
    (Left) The difference measure is SNR (dB) between the original signal and the filtered signals by 10000 unit-CAPRICEPs.
    (Right) Effects on sampled waveform value distribution. The black line is the original. The gray 2000 lines are filtered ones.}
    \label{fig:augmentation}
\end{figure}
The end-to-end approach requires an enormous amount of training data\cite{DLforAudio}.
When data is scarce, such as low-resource languages, data augmentation applies.
Using many unit-CAPRICEPs provides arbitrarily many perceptually identical samples from a single original sound material.
Figure~\ref{fig:augmentation} shows two examples.
Left uses music data excerpted from\cite{goto2002rwc}.
Despite this low-SNR distribution, the impression of the filtered sounds
is identical to the original.
No impression of noise.
%The plot shows the distribution of the difference between the original signal and augmented samples, measured in terms of SNR.
Right shows a voice (Japanese vowel /e/) example\cite{nakayama2004ica}.
Filtering by unit-CAPRICEPs modifies the waveform. 
The original asymmetric distribution (thick black lines) turned into less asymmetric distribution (gray 2000 lines) by filtering.
The filtered signals sound perceptually identical despite that they are significantly different in the waveform.
Sample files are in the first author's GitHub repository~\cite{kawahara2021capricep}.

TSP-based methods\cite{schroeder1979integrated,aoshima1981jasa,muller2001transfer,ochiai2013a,guidorzi2015impulse} are commonly used in acoustic measurement and appropriate for measuring linear time-invariant responses.
However, they need extra-equipment or inspection by experts for analyzing non-linear responses and inter-modulation distortion\cite{dunn1993distortion,farina2000simultaneous,stan2002comparison,burrascano2019swept}.
FVN solved this problem by proposing a simultaneous measurement method that does not require extra-equipment nor experts' inspection\cite{kawahara2020simultaneous}.
The proposed CAPRICEP revised the FVN-based method with a theoretically solid and more flexible foundation.
The proposed method is general enough for investigating systems with inherent non-linear and temporal variability, biological systems.
We are planning to apply the CAPRICEP-based method for investigating the auditory-to-speech production mechanism\cite{kawahara1994interactions,Kawahara1996is,TOURVILLE20081429,houde2013PNAS,Zarate2013front}.

\vspace{-4pt}
\section{conclusion}
\vspace{-2pt}
We proposed a new member of TSP called CAPRICEP.
A unit-CAPRICEP is an impulse response of an all-pass filter with high degrees of freedom in design and provides a theoretically sound infrastructure for acoustic measurement and assessment.
CAPRICEP opens vast application possibilities spanning from acoustic measurement, speech synthesis, signal processing for biological systems, and data augmentation for the end-to-end approach of sound-related applications.
It also provides an essential key for fundamental research of human auditory perception principles.
We made the CAPRICEP-based interactive and real-time acoustic measurement tools and related MATLAB functions open-source in the first author's GitHub repository\cite{kawahara2021capricep}.

\bibliographystyle{IEEEbib}
\bibliography{kawaharaArXiV}

\end{document}